\documentclass{PoS}
\title{Recent Progress on the GAPS Time of Flight System}

\ShortTitle{GAPS TOF}

\author{\speaker{S. Quinn}$^{,1}$ for the GAPS collaboration$^{a,}$\footnote{for collaboration list see PoS(ICRC2019)1177}\\
	\llap{$^1$}Deptartment of Physics \& Astronomy, University of California, Los Angeles, CA, USA\\
        E-mail: \email{spq@ucla.edu} \\
$^{a}$Full author list: \href{https://gaps1.astro.ucla.edu/gaps/authors/alist.html}{\rm gaps1.astro.ucla.edu/gaps/authors/alist.html}
}

\abstract{
The General AntiParticle Spectrometer (GAPS) is a balloon-borne cosmic-ray antimatter experiment that uses the exotic atom technique, eliminating the requirement for strong B-fields used by conventional magnetic spectrometers. It will be sensitive to antideuterons with kinetic energies of 0.05-0.25 GeV / nucleon, which are highly motivated candidates for indirect dark matter detection. Moreover, GAPS will provide new information on the antiproton spectrum from $0.07<T<0.25$ GeV. The GAPS design is based on a lithium drifted silicon tracker and plastic scintillator time of flight (TOF) system. The latter is the focus of this contribution.
	
Currently, the TOF system includes an outer ``umbrella'' consisting of 132 counters covering an area of 38 m$^2$ and a nearly hermetic inner ``cube'' with 64 counters and area of 15 m$^2$. The counters will be mechanically secured to the gondola using an innovative carbon fiber structure. Each end of the 196 counters will be read out using a silicon photomultiplier (SiPM) based analog front end with a high gain timing channel and low gain trigger channel. The high gain channel is sampled and digitized with a custom readout board that uses the DRS-4 ASIC. A local trigger monitors multiple programmable threshold levels for all 392 counter ends. A master trigger analyzes the local trigger hit patterns and initiates a TOF read out for an interesting event. A central computer then analyzes and estimates key observables. This contribution summarizes the design, performance, and prototype development of the TOF system and the path going forward in 2019 and 2020 towards construction and integration of the system.
}

\FullConference{36th International Cosmic Ray Conference -ICRC2019-\\
		July 24th - August 1st, 2019\\
		Madison, WI, U.S.A.}

\begin{document}

\section{Introduction}

GAPS is a balloon-borne experiment designed to detect light, non-relativistic cosmic ray antimatter. The primary goal is to shed new light on the low kinetic energy antiproton/antideuteron spectrum, which provides complementary data to existing experiments at higher energies. Beyond some usefulness in constraining cosmic ray propagation models, low-energy antimatter serves as exciting indirect probes for dark matter---$\overline{d}$ in particular, whose primary dark matter flux exceeds the expected background astrophysical flux by $\approx100$. Another major goal is sensitivity to antihelium. For a brief review of the particle physics motivation for GAPS see \cite{QuinnCIPANP18}, and for a detailed science case see \cite{Physrep2016}. The current iteration of GAPS builds on the heritage of the successful prototype 2012 flight \cite{pGAPSNIMA}.

\section{Detector overview}
Instead of relying on track bending used by a traditional magnetic spectrometer, GAPS employs the exotic atom technique integrated with a large area time of flight. Antimatter primaries entering the instrument are degraded by active and dead material until they are stopped, forming an unstable exotic atom. As this atom decays, it generates characteristic X-rays dependent on the primary species. Moreover, the multiplicity of pions and other annihilation products provide additional handles for primary identification. For a more detailed description of the detection technique and a general detector overview see \cite{GAPSICRC2019}.

The main target of GAPS consists of multiple layers of Si(Li) wafers used for vertex finding, particle tracking, and high energy X-ray detection. For a detailed description of the fabrication and characterization of these detectors see \cite{SiLiNIMA2018}. A custom ASIC \cite{SiLiASICIEEE2019} and analog front end \cite{SiLiFrontEnd} are being developed for data acquisition of the 11,520 input channels.

\begin{figure}
	\begin{center}
	\includegraphics[scale=0.13]{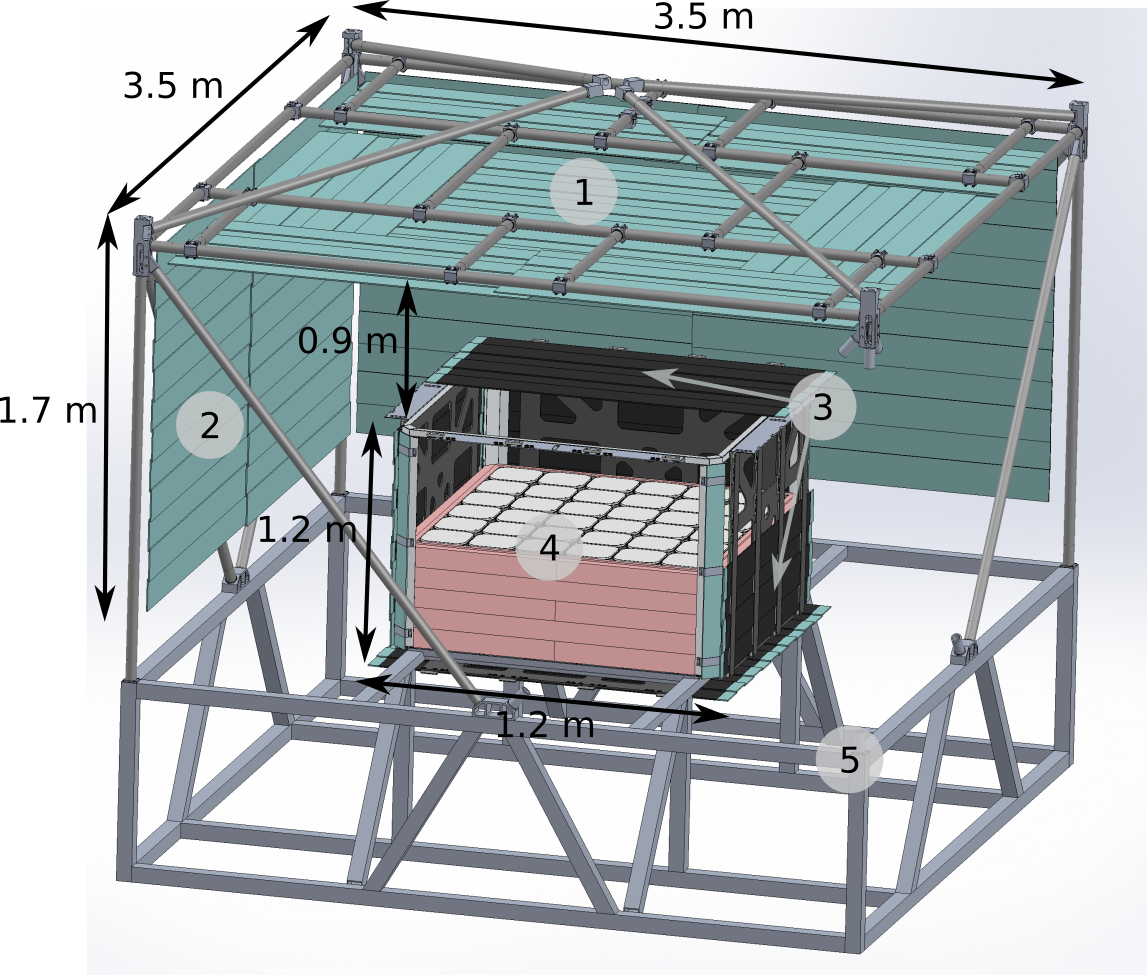}
	\end{center}
	\caption{\label{FIG:cutaway}Cutaway CAD model of current GAPS payload with key dimensions. (1) umbrella top (2) umbrella cortina (3) cube, sealed with black wrap and also showing exposed carbon fiber skeleton (4) Si(Li) tracker and (5) gondola frame.}
\end{figure}

The Si(Li) tracker is surrounded by a time of flight (TOF) system using plastic scintillators in a two-layer design. The ``umbrella'' layer has a top area which can loosely be taken to define the instrument bore for downward going particles. The umbrella sides (``cortina'') expose an effective area for more inclined primaries. The second layer is a ``cube'' which completely envelops the Si(Li) tracker system. With this design the TOF is able to (1) calculate primary $\beta$, (2) providing additional tracking information for annihilation tracks and an important time stamp to discriminate between in and outgoing particles, (3) estimate primary composition, and (4) supply the trigger to read out the Si(Li) system. A rendering of the instrument is shown in Figure \ref{FIG:cutaway}.

%Goals of TOF system: $\beta$ measurement, particle id, particle position. Overview of scintillation physics.

\section{Time of flight overview}
At its core, the TOF relies on the proven technology of plastic scintillators for relativistic charged particle detection. The measurement chain begins with the production of isotropic near UV scintillation photons which propagate via total internal reflection to the counter ends, into an optically coupled low-light sensor. A custom analog front end (preamp) then outputs measurable electronic signals into several downstream custom boards used to make trigger decisions, as well as to digitize waveforms. In this section, the design and current performance of these subsystems will be described in further detail, starting with the mechanical layout.

\section{Mechanical design}
The two TOF layers are composed of individual counters, or paddles, of Eljen EJ-200 plastic scintillator of thickness 6.35 mm and width 16 cm. The umbrella uses 1.8 m lengths, while the cube uses 1.8 m, 1.56 m, and 1.1 m lengths. Since counters are overlapping in both layers, the faces will have 100 \% hermeticity. In some regions, due to mechanical constraints, the worst value of the hermeticity will be $\approx96$ \%. Counters will be strapped to carbon fiber panels which are then mounted to the gondola frame. The overall structural integrity is designed to withstand large impacts for launch and recovery.

\begin{figure}
	\begin{center}
	\includegraphics[scale=0.115]{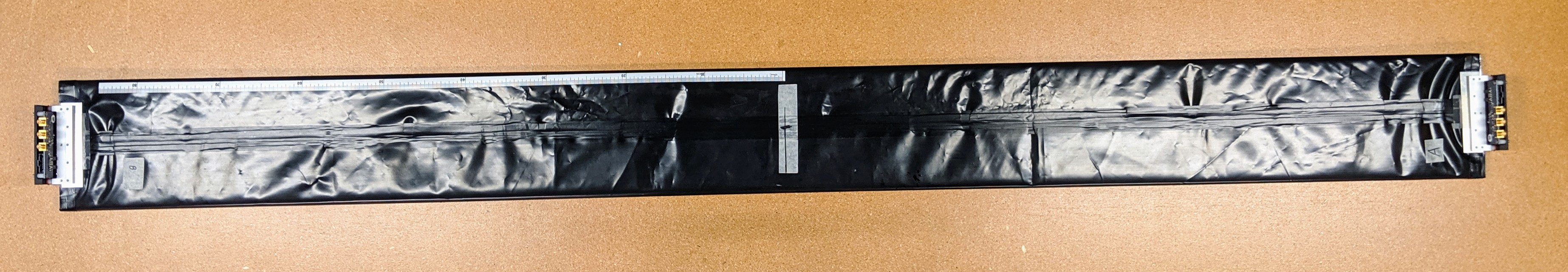}
	\end{center}
	\caption{\label{FIG:padimg}Prototype counter. The preamps have been mounted, and the main body sealed. This preamp mounting variant is not light tight, so a dark box is used in lab testing.}
\end{figure}

The machined scintillator is first wrapped with a base layer of aluminum foil, which gets overwrapped and sealed with opaque black plastic (Delta Blackout). A custom U-channel assembly and preamp enclosure are secured directly to the counter. A partially complete 1.8 m type is shown in Figure \ref{FIG:padimg}. For lab testing, Eljen EJ-550 optical grade silicone grease is used for optical coupling. For the flight version, we are investigating a variety of flexible adhesive strategies. The following sections describe the signal path and data acquisition.

\section{Analog front end}
After an extensive comparison between Hamamatsu R7600-200 photomultiplier tubes (PMT), Hamamatsu S14160-6050HS and S13360-6050VE silicon photomultipliers (SiPMs) the TOF system has adopted the S13360-6050VE model. In general SiPMs, are more cost-effective and their small form factor permits coupling directly to the polished scintillator. This obviates the need for bulky light guides required by PMTs and provides a significant margin for the overall mass budget.

The \href{https://www.hamamatsu.com/us/en/product/type/S13360-6050VE/index.html}{S13360-6050VE} is a 6x6 mm surface mount SiPM with excellent noise characteristics and low terminal capacitance. To meet demanding timing requirements, we adopt a nominal operating gain of $G=4\times10^6$, corresponding to $V_{bias}=58.6$ V at room temperature.

\begin{figure}
	\begin{center}
	\includegraphics[scale=0.182]{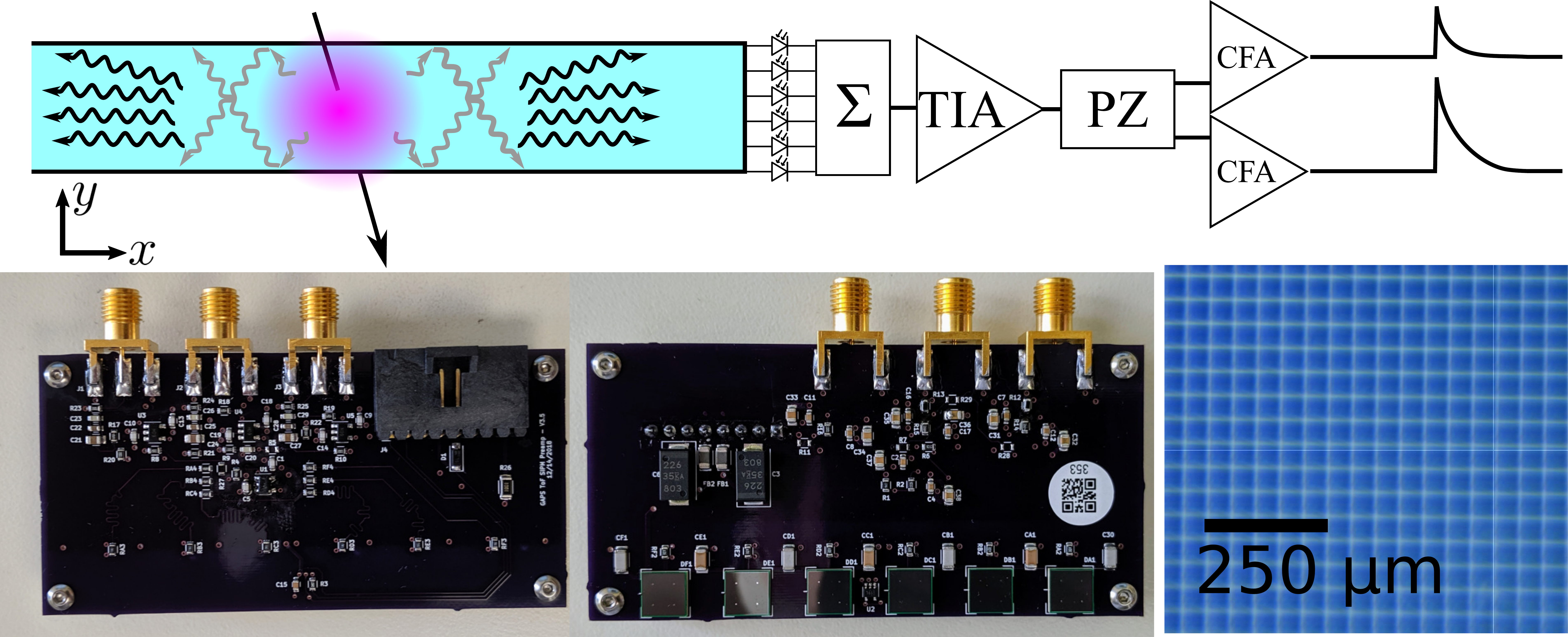}
	\end{center}
	\caption{\label{FIG:preamp}\textit{Top:} Simplified cartoon of the detection process and analog front end signal path. Black photons are prompt light contributing to the fast rising edge, while the gray delayed photons arrive shortly after. \textit{Bottom left:}  View of preamp board underside showing power and sense connector and op-amps. \textit{Bottom middle:}  Top view of preamp board with SiPMs and various bypassing networks. \textit{Bottom right:} Magnified view of SiPM package showing individual avalanche photodiode cells.}
\end{figure}

The TOF preamp design uses a three-stage op-amp based architecture. The first is a balanced analog sum of 6 SiPMs followed by a transimpedance amplifier with a nominal amplitude gain of $G_V=19.5$. Lastly, a pole zero cancellation splits the output into two current feedback stages. One is configured as a high gain timing channel with $G_V=9.6$, and the second is a low gain trigger channel with $G_V=1.8$. A high-level block diagram summarizes the design in Figure \ref{FIG:preamp}.

The front end is supplied with 5V for op-amps, and a fully programmable high voltage rail is used for the SiPMs. Since breakdown voltage is known to vary with temperature \cite{OtteNIMA}, a central temperature sensor adjacent to the SiPMs is actively monitored, and the high voltage is adjusted in real time. A V3.5 prototype board is shown in Figure \ref{FIG:preamp}. Preamp outputs are then sent over a 50 $\Omega$ low-loss transmission line into a data acquisition (DAQ) system.

\section{Custom low power DAQ}
The scintillator waveform is rich with information about the transiting particle, so we are proceeding with a design to sample and digitize all counters which participate in a physics trigger. This approach enables further analysis using powerful machine learning techniques in software which would be challenging to implement in a purely hardware-based architecture. While this approach offers many advantages, it must also conform to demanding power and density requirements.

\subsection{DRS4}
To achieve high timing resolution with minimal power consumption, the DRS4 switched capacitor array ASIC (which can be thought of as a type of analog memory) is used \cite{DRS4NIMA}. The device offers low timing jitter ($\approx 100$ ps), nine high bandwidth analog inputs, 1024 samples at 2 GS/s using 140 mW of power in a small package. In our design, the nine analog outputs will be multiplexed into a 14-bit ADC which is further processed by an FPGA.

For optimal performance, an amplitude and timing calibration with well-characterized sources is done to understand the behavior of the 1024 sampling cells. The TOF readout board will include a precision DAC and temperature compensated crystal oscillator for \textit{in-situ} calibration runs. The analysis and calibration corrections will be determined using the methodology given in \cite{DRS4CAL}.

\subsection{V2 readout board design}
The prototyping phase for this project began in early 2018 with the fabrication of the V1 readout board. The DRS4 evaluation board offered by PSI is used for lab testing. As a result, we adopted the analog input architecture (same differential op-amps and ADC) used on that board to maintain consistency. To improve density, the V1 board used two DRS4 chips for a total of 16 inputs. For firmware compatibility, a legacy Spartan 3 FPGA was used which included a custom SPI interface to control the DRS4 and peripherals. In testing, basic functionality was demonstrated, but signal quality issues were observed.

\begin{figure}
	\begin{center}
	\includegraphics[scale=0.5]{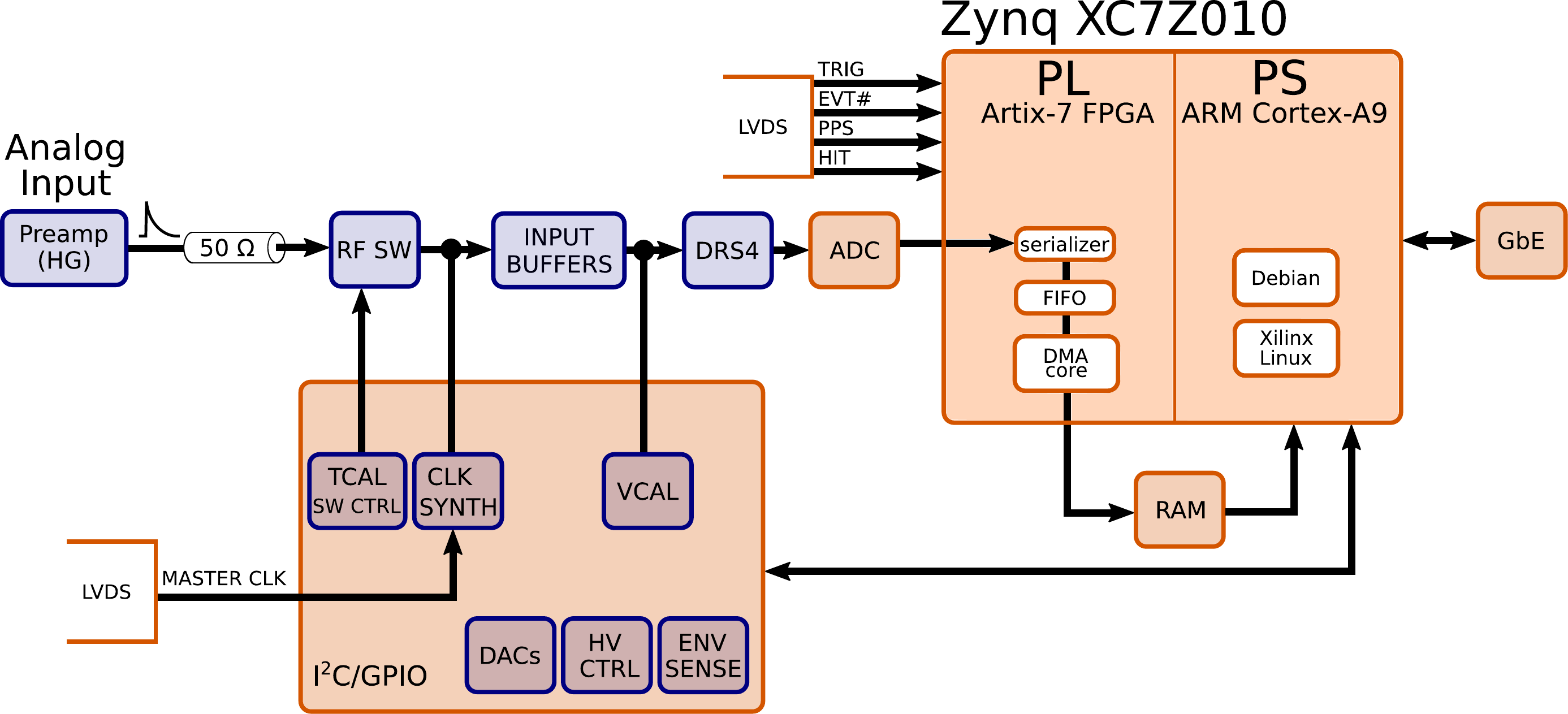}
	\end{center}
	\caption{\label{FIG:roblock}High level overview of the data acquisition system for V2 readout board. Analog and sensing components appear as blue boxes while digital components are orange. Only one input path, preamp high gain (HG), is drawn for clarity, but it is duplicated seven times in the actual design. Timing calibration (TCAL) is done using special purpose switches (SW). Global clock synchronization (CLK SYNTH) is achieved by sampling a continuously running sine wave into the last DRS4 input channel.}
\end{figure}
\begin{figure}
	\begin{center}
		\includegraphics[scale=0.78]{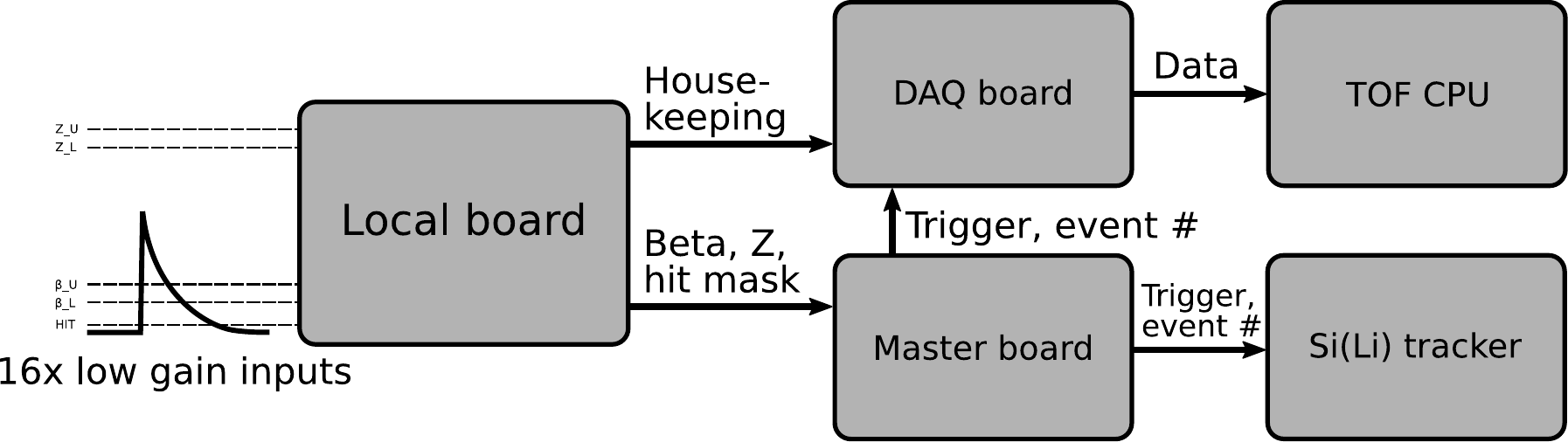}
		\caption{\label{FIG:trigblock}The main, high level elements of the trigger design. Charge (Z) levels are used to screen out heavy nuclei and are expected to be $\sim10$ larger than the velocity ($\beta$) levels which help reject fast light nuclei.}
	\end{center}
\end{figure}

Design work for the second iteration began in fall 2018 and improves on V1 in several key areas: (1) shortening the distance between input buffers, ADC and DRS4 to avoid distributed elements, (2) implementing a Mars Enclustra ZX2 Zynq-7000 based system on module (SOM) to simplify fast data handling (via direct memory access) as well as slow control peripherals, (3) incorporating a low jitter clock synthesizer for global synchronization and absolute timing capabilities, (4) Gigabit Ethernet connectivity to simplify data transfer over long distances and (5) isolation of digital and analog supplies along with galvanic isolation of inputs to prevent ground loops. A block diagram highlighting the main subsystems is shown in Figure \ref{FIG:roblock}. We expect to fabricate and begin testing V2 prototypes by late summer 2019.

\section{Trigger}
The trigger consists of a two-level hierarchical design. The first level, the ``local trigger'', monitors 16 channels (8 counters) and uses fast discriminators with five adjustable thresholds to identify if a channel has been hit and whether that signal is consistent with slow ($\beta\lesssim0.6$), light ($Z=1$ or 2) candidates. The local trigger digital outputs are streamed to a master board which searches for a slow, light primary track (umbrella hit followed by cube hit) associated with multiple channels hit by fast ($\beta\sim1$) particles to identify candidate antinuclei annihilation stars. Though primarily tuned to search for antideuterons, the trigger requirements will also have good sensitivity to antiprotons and heavier antinuclei while providing a high level of rejection to other cosmic rays.  When coincidence, hit pattern, and level conditions are satisfied, a trigger is asserted to all TOF data acquisition boards, and the Si(Li) tracker system for event read out. An offline trigger level will be employed to identify potentially interesting events for saving to disk.  A notional diagram of the functionality and data flow is shown in Figure \ref{FIG:trigblock}.

\section{Current performance}
\begin{figure}
	\begin{center}
	\includegraphics[scale=0.1]{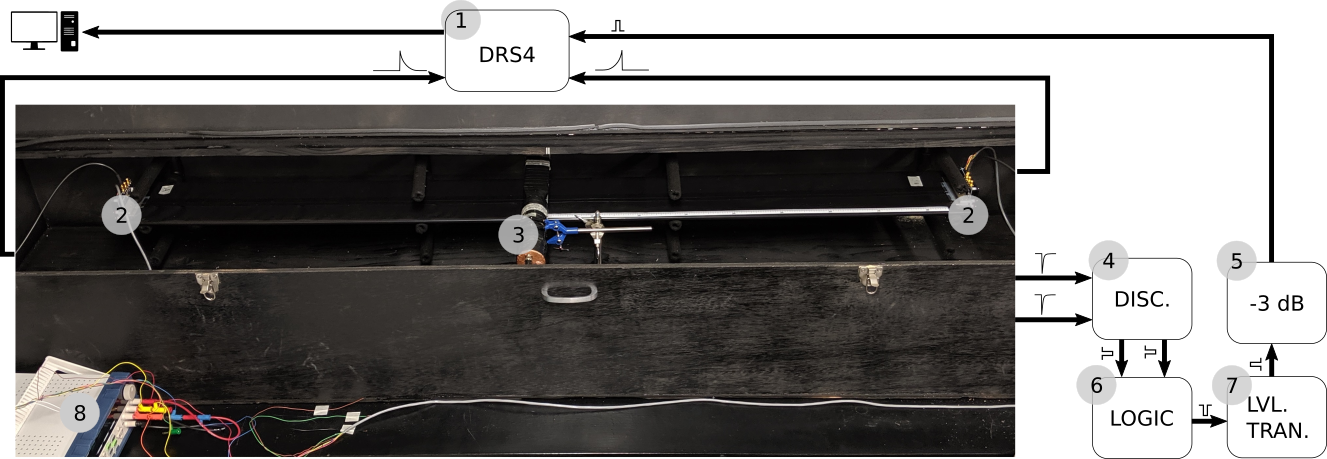}
	\end{center}
	\caption{\label{FIG:darkbox}Dark box test stand. (1) V5.1 DRS4 evaluation board, (2) UCLA V3.5.3 lab variant preamp, (3) hodoscope with 14 cm separation, (4) Phillips Scientific 711 discriminator, (5) -3 dB RF attenuator, (6) Phillips Scientific 755 quad four fold logic unit, (7) Phillips Scientific 711 NIM to TTL translator and (8) BK  Precision 9132B supply for preamp op-amps and SiPM high voltage.}
\end{figure}

Here we describe the lab test stand used for benchmarking prototypes. The outputs of two small scintillators with gain matched PMTs are sent into NIM discriminators and coincidence logic (40 ns gate) modules to form a trigger on atmospheric muons (zenith angle $<35^{\circ}$). For a photo of the setup, see Figure \ref{FIG:darkbox}.

\subsection{Timing}
\begin{figure}
	\includegraphics[scale=0.39]{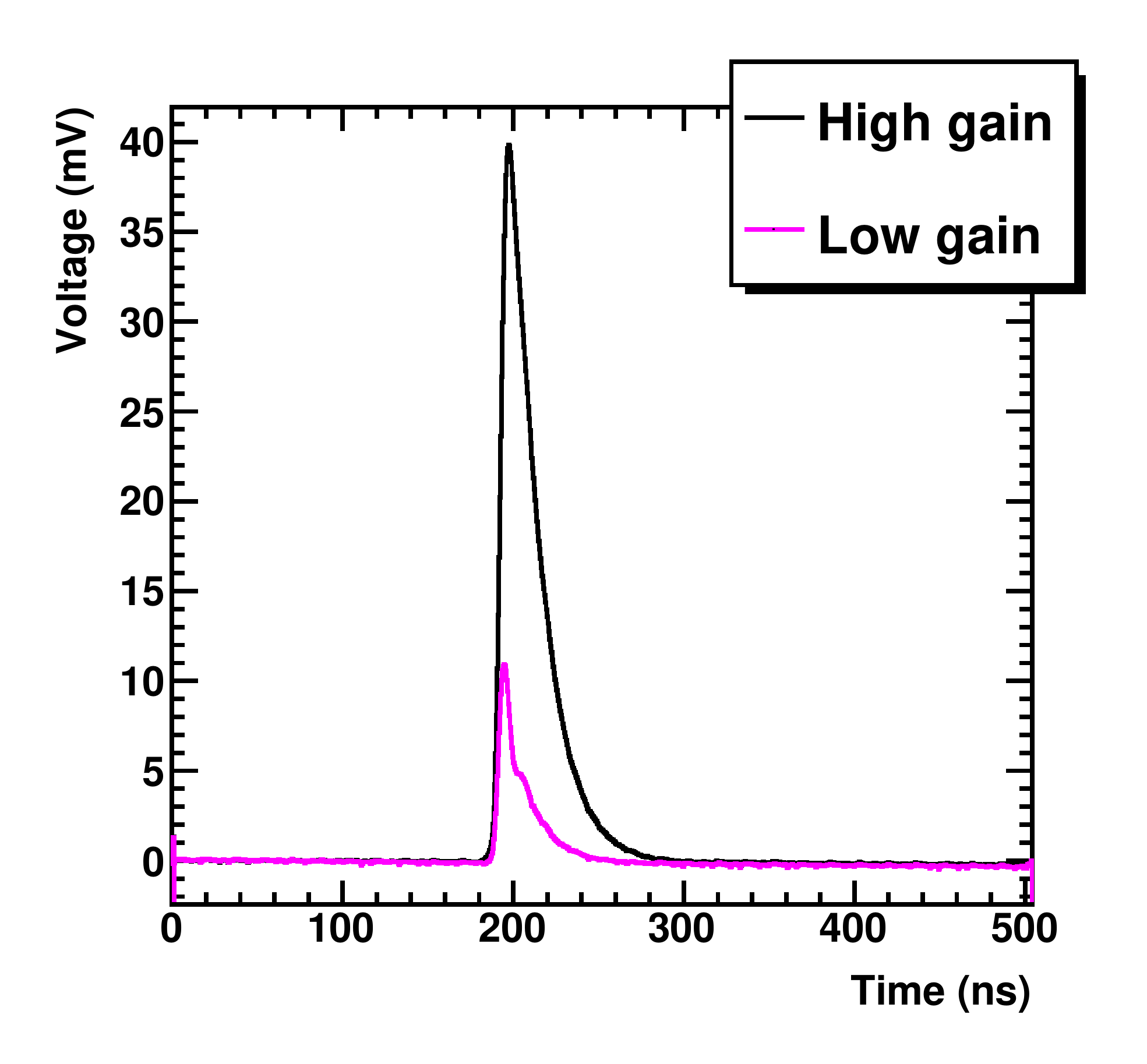}
	\includegraphics[scale=0.37]{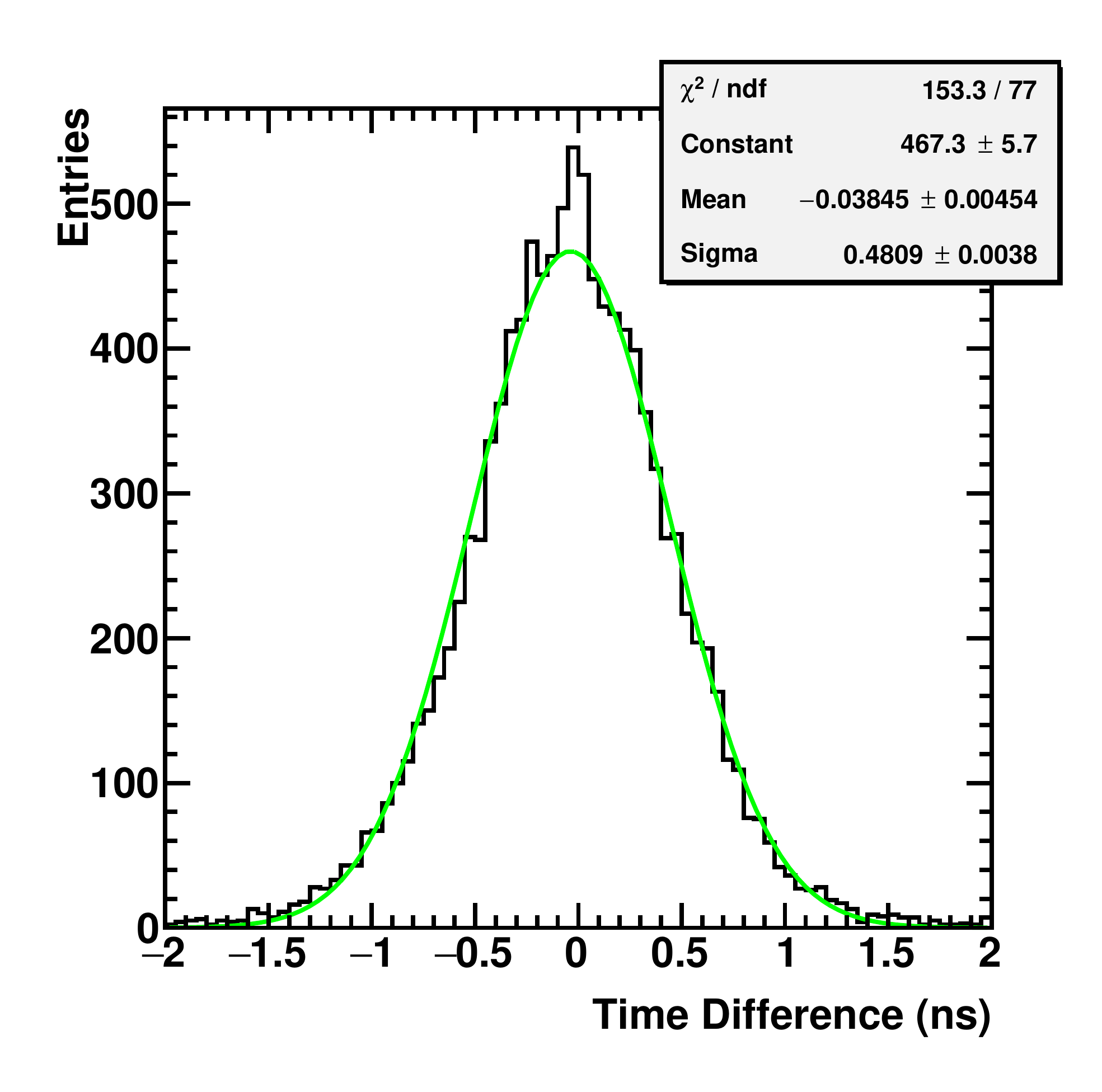}
	\caption{\label{FIG:trace_timing}\textit{Left:} Mean waveforms for central vertical muons. The HG pulse will be used for $\beta$ calculation while the LG pulse is used for triggering and veto of heavy nuclei. \textit{Right:} $t_A-t_B$ histogram with a Gaussian fit.}
\end{figure}

A key requirement for the TOF system is achieving a 500 ps timing resolution ($\delta t$). This requirement has a direct impact on the resolution in determining the velocity ($\beta$). It also impacts the position reconstruction of the track (in the form of uncertainty in the long, termed $x$, direction along the counter). As $\delta t$ decreases, so does $\delta x$. In the $y$ direction the resolution is bounded by the narrow dimension (16 cm). To estimate $\delta t$, the time difference between the counter ends is calculated for digitized data using a software constant fraction discriminator (CFD). The timing resolution is defined as $\sigma \left( t_A - t_B\right) / \sqrt{2}$ where $t_X$ is the threshold crossing time at either end. High gain timing results ($\delta t=340$ ps) for a typical run with $N=11,577$ events are presented in Figure \ref{FIG:trace_timing}.

\section{Conclusion}
To meet the GAPS mission requirements, a novel carbon fiber based mechanical support structure is used for the 196 counters. The unique geometry, environmental conditions, and rapid timeline combine to form a challenging design target.  A dual output preamp for high precision timing and high dynamic range is sampled with a low power custom DRS4 based DAQ board. The TOF will be used to form a physics trigger for the GAPS instrument using a hierarchical scheme based on a fast local trigger with programmable thresholds and powerful master trigger that analyzes hit patterns. In this proceeding, we have shown the latest results for the TOF timing resolution in addition to the improvements and the new architecture of the V2 readout board along with the prototype trigger system. We intend to continue to progress over the summer during the TOF pre-production in which 10\% of the TOF system will be commissioned and characterized.


\begin{thebibliography}{99}
	\bibitem{QuinnCIPANP18}GAPS Collaboration, \href{https://arxiv.org/abs/1809.09714}{Proceedings of 13th Conference on the Intersections of Particle and Nuclear Physics} (CIPANP 2018) %1
	\bibitem{Physrep2016}T.~Aramaki et al., \href{https://doi.org/10.1016/j.physrep.2016.01.002}{Physics Reports 618 1-37 (2016)}  %2
	\bibitem{pGAPSNIMA}S.A.I.~Mognet et al., \href{https://doi.org/10.1016/j.nima.2013.08.030}{NIMA 735 24-38 (2014)}   %3
	\bibitem{GAPSICRC2019}R.~Bird, These Proceedings %4	
	\bibitem{SiLiNIMA2018}K.~Perez et al., \href{https://doi.org/10.1016/j.nima.2018.07.024}{NIMA 905 12-21 (2018)}
	\bibitem{SiLiASICIEEE2019}M.~Manghisoni et al., In preparation
	\bibitem{SiLiFrontEnd}V.~Scotti, These Proceedings
	\bibitem{OtteNIMA}A.N.~Otte et al., \href{https://doi.org/10.1016/j.nima.2016.09.053}{NIMA 846 106-125 (2017)}
	\bibitem{DRS4NIMA}S.~Ritt, \href{https://doi.org/10.1016/j.nima.2003.11.059}{NIMA 518 470-71 (2004)}
	\bibitem{DRS4CAL}D.~Stricker-Shaver, S.~Ritt, B.J.~Pichler, \href{https://doi.org/10.1109/TNS.2014.2366071}{IEEE Transactions on Nuclear Science 61 3607-3617 (2014)} 
	
\end{thebibliography}
\end{document}